\documentclass{article}

\usepackage{arxiv}

\usepackage[utf8]{inputenc} 
\usepackage[T1]{fontenc}    
\usepackage{hyperref}       
\usepackage{url}            
\usepackage{booktabs}   
\usepackage{cite} 
\usepackage{amsfonts}       
\usepackage{nicefrac}       
\usepackage{microtype}      
\usepackage{lipsum}
\usepackage{graphicx}

\usepackage{balance}       
\usepackage{graphics}      
\usepackage[T1]{fontenc}   
\usepackage{txfonts}
\usepackage{wrapfig}
\usepackage{booktabs}
\usepackage{textcomp}
\usepackage{tabularx}
\usepackage{soul}
\usepackage{url}
\usepackage{multirow}
\usepackage{multicol}
\usepackage{xcolor}
\graphicspath{ {./images/} }

\definecolor{Orange}{rgb}{1,0.5,0}
\definecolor{Red}{rgb}{1,0,0}
\definecolor{Green}{rgb}{0,0.8,0.5}
\definecolor{Purple}{rgb}{0.75,0,1}
\definecolor{babypink}{rgb}{0.96, 0.76, 0.76}
\definecolor{azure}{rgb}{0,0.49,1}
\definecolor{periwinkle}{rgb}{0.8, 0.8, 1.0}
\definecolor{Pink}{RGB}{255, 102, 204}

\title{How good is good enough for COVID19 apps?\\
\Large{The influence of benefits, accuracy, and privacy on willingness to adopt}}

\author{
  Gabriel Kaptchuk \\
  Johns Hopkins University \\
   \texttt{gkaptchuk@jhu.edu} \\
  \And
  Daniel G. Goldstein \\
  Microsoft Research\\
  \texttt{dgg@microsoft.com}\\
  \And
 Eszter Hargittai \\
 University of Zurich \\
  \texttt{pubs@webuse.org} \\
   \And
   Jake M. Hofman\\
   Microsoft Research\\
   \texttt{jmh@microsoft.com}\\
   \And
   Elissa M. Redmiles \\
 Microsoft Research\\
  \texttt{Elissa.Redmiles@microsoft.com} 
}

\begin{document}

\maketitle
\begin{abstract}
A growing number of contact tracing apps are being developed to complement manual contact tracing. A key question is whether users will be willing to adopt these contact tracing apps. In this work, we survey over 4,500 Americans to evaluate (1) the effect of both accuracy and privacy concerns on reported willingness to install COVID19 contact tracing apps and (2) how different groups of users weight accuracy vs. privacy. Drawing on our findings from these first two research questions, we (3) empirically model how the \textit{amount} of public health benefit (reduction in infection rate), \textit{amount} of individual benefit (true-positive detection of exposures to COVID), and \textit{degree} of privacy risk in a hypothetical contact tracing app may influence Americans' willingness to install. Our work takes a \textit{descriptive ethics} approach toward offering implications for the development of policy and app designs related to COVID19.
\end{abstract}
\section{Introduction}
A growing number of coronavirus (COVID19) contact tracing apps are being developed and released with the goal of tracking and reducing the spread of COVID19~\cite{AScrambl95:online}. These apps are designed to complement manual contact tracing efforts using location data or Bluetooth communication to automatically detect if a user may have been exposed to the virus~\cite{chan2020pact, troncoso2020decentralized}.  Unlike manual contact tracing, which an investigator reaches out directly to affected parties, the benefits of these apps for public health scale {\em quadratically} with participation~\cite{chan2020pact}.  This is because both data collection and data distribution are part of the apps operation.  Thus, it is critical to understand that factors that determine if people will be willing to adopt these apps.

There are a large number of considerations that may influence users willingness to adopt~\cite{redmiles2020user}.  For example, a person may weigh the features the app's offer, the app's benefits to themselves and their community~\cite{simko2020}, the provider offering the app~\cite{WillAmer35:online}, how well the app will preserve the user's privacy~\cite{troncoso2020decentralized,chan2020pact}, and the app's accuracy.  Understanding the impact of each of these factors can help app developers make design decisions that can maximize their impact. 

Drawing from the idea of descriptive ethics as a more fair approach to setting societal norms~\cite{grgic2018human,gert2002definition}, in this work we use surveys to evaluate {\em how well} COVID19 apps need to function for users to be willing to adopt the apps. We present the results of a series of surveys of a total of 4,615 Americans, sampled using both crowd-sourcing and online survey panels that can satisfy census-representative demographic quotas. 

There are many ways to measure how well a COVID19 app works. From a public health perspective, reducing infection rate (i.e., basic reproduction number) is a key measurement of success. However, in order to understand individual's choices when it comes to adopting an app, we must also consider what it means for a COVID19 app to work well for {\em the individual} who has the app installed. 

Thus, we examine not only the societal-level, public health benefit of infection rate reduction, but also how app accuracy and app privacy, risk of the app exposing information collected by the app to others, influence reported willingness to adopt a COVID19 app.  Within app accuracy, we consider both false negatives, the app failing to detect an exposure to COVID19, and false positives, the app falsely notifying the user that they were exposed when they were not. Understanding how rates of app failure may influence adoption allow us to estimate user response to potential app designs. For example, Saxena et. al. ~\cite{Smartpho90:online} have shown that using Bluetooth as a method for detecting proximity may be innately error prone, estimating an approximate error rate of 7-15\% (including both false positive and false negative rates). Our results allow us to estimate user response to such error rates.

In summary, we address the following three research questions: 
\begin{enumerate}
\item [\textbf{(RQ1)}] Do both accuracy (precision and recall) \& privacy influence whether people want to install a COVID app? 
\medskip
\item [\textbf{(RQ2)}]   Do different types of people weight accuracy or privacy more heavily?
\medskip
\item [\textbf{(RQ3)}]   How much public health benefit, accuracy, and/or privacy is necessary for people to want to adopt COVID19 apps? 
\end{enumerate}

%

We find that:
\begin{itemize}
    \item Between 70-80\% of Americans report being willing to install an app that is ``perfectly'' private and/or accurate, a significant increase from the 50-60\% who are willing to install an app with unspecified privacy or accuracy~\cite{WillAmer35:online, Washingt8:online}. 
    \item False negatives have a significantly stronger influence on reported willingness than false positives or privacy risks. 
    \item Reported willingness to install correlates  with the public health benefit and/or personal health benefit of a contact tracing app. Specifically, the majority of Americans report being willing to install an app that offers at least a 50\% improvement in public health or in personal safety over the baseline rate offered when not using the app.
\end{itemize}
%

\section{Methodology}
We conducted a series of surveys to answer our research questions. In this section we discuss our questionnaires, questionnaire validation, sampling approaches, analysis approaches, and the limitations of our work. All studies were IRB approved by the Microsoft Research IRB, a federally recognized ethics review board.

\subsection{RQ1 and RQ2 Survey}
\label{sec:s1methods}
In this first survey we sought to understand how accuracy and/or privacy considerations might influence willingness to adopt (RQ1) and how respondent demographics and experiences might affect the relative weight of these considerations (RQ2). 

We used a vignette survey~\cite{atzmuller2010experimental} to examine these questions as vignette surveys are known to maximize external validity. 

\textbf{Questionnaire.} Our questions were framed around a contact-tracing app scenario. Half of the respondents were placed in the proximity contact tracing scenario while the other half were placed in the location scenario. 

The proximity scenario was phrased as: 
\begin{quote}
    Imagine that there is a mobile phone app intended to help combat the coronavirus. This app will collect information about who you have been near (within 6 feet), without revealing their identities.\\

The app will use this information to alert you if you have been near someone who was diagnosed with coronavirus. If you decide to inform the app that you have been diagnosed with coronavirus, the app will inform those you've been near that they are at risk, without revealing your identity.
\end{quote}

The location scenario was phrased as:

\begin{quote}
    Imagine that there is a mobile phone app intended to help combat the coronavirus. This app will collect information about your location.\\
    
    The app will use this information to notify you, without revealing anyone's identity:
\begin{itemize} 
\item if you have been near someone who tested positive for coronavirus 
\item about locations near you that were recently visited by people who tested positive for coronavirus\\
 \end{itemize}
 
If you decide to report to the app that you have been diagnosed with coronavirus, the app will inform those you've been near that they are at risk without revealing your identity.
\end{quote}

Participants were then routed to a set of control questions or a set of experimental questions regarding accruacy and privacy (in randomized order). All participants were asked ``Would you install this app?'' after a given question, with answer choices ``Yes'', ``No'', and ``It depends on the [risk, chance of information being revealed, etc.]''.

\textit{Control.} We had three control conditions (respondents saw only one of these three questions).

Perfect accuracy:
\begin{quote}
    Imagine that this app will work perfectly.\\ 

It will never fail to notify you when you are at risk nor will it ever incorrectly notify you when you are not at risk.
\end{quote}

Perfect privacy:
\begin{quote}
    Imagine that this app perfectly protects your privacy.\\ 

It will never reveal any information about you to the US government, to a tech company, to your employer, or to anyone else.
\end{quote}

Perfect accuracy and privacy:
\begin{quote}
    Imagine that this app works perfectly and protects your privacy perfectly.\\

It will never fail to notify you when you are at risk nor will it ever incorrectly notify you when you are not at risk. It will also never reveal any information about you to the US government, to a tech company, to your employer, or to anyone else.
\end{quote}

\textit{Experimental.}
These participants were asked about accuracy and privacy, in randomized order. 

Accuracy (false negatives):
\begin{quote}
    Imagine that this app occasionally fails to notify you when you have been near someone who was diagnosed with coronavirus.
\end{quote}

Accuracy (false positives): 
\begin{quote}
    Imagine that this app occasionally notifies you that you have been near someone who has coronavirus when you actually have not been exposed.
\end{quote}

Privacy:
\begin{quote}
    Imagine that this app might reveal information about [who you have been near/your location] to [entity]. This information may be used for a purpose unrelated to the fight against coronavirus.
\end{quote}
We asked about four entities, drawn from the list of 10 examined by Redmiles and Hargittai~\cite{WillAmer35:online}: ``non-profit institutions verified by the government'', ``technology companies'', ``the US government'', and ``your employer''.

\textbf{Validation.} The questionnaire design was validated through expert reviews with multiple external researchers. Additionally, three attention check questions were included, one general attention check and two scenario-specific attention checks that ensured respondents understood the scenarios described. 

\textbf{Sample.} 789 Americans answered our survey. The sample was quota sampled by Cint to be representative of the US population demographics on age, gender, income, and race. 

\textbf{Analysis.} We answered RQ1 using $X^2$ proportion tests to compare responses to our different sets of questions. We answered RQ2 by constructing two mixed effects binomial logistic regression models. In both models, our dependent variable was willingness to install the app, with ``Yes'' and ``It depends on the risk'' grouped together as a positive outcome and ``No'' was treated as a negative outcome. We model responses to the accuracy and privacy questions separately, controlling for data type and entity, in the privacy model, and both data and accuracy type in the accuracy model. We included as dependent variables the respondents' age, gender, race, internet skill (as measured using the Web Use Skill Index~\cite{hargittai2009update}), level of educational attainment, party affiliation, and if they know someone who died due to complications from COVID19.  Finally, we include a mixed effects term to account for our within subjects design. 

\subsection{RQ3 Survey}
In this survey we sought to evaluate how people's reported willingness to install coronavirus apps correlates with the \textit{amount} of public health (infection rate reduction) and individual health (notification of at risk status -- e.g., accuracy) benefit of a hypothetical coronavirus tracking app.

\textbf{Questionnaire.} All questions, except one control condition (FN app control, addressed below), were asked in the context of the following scenario. As the \textit{type} of information compromised, as well as the entity that could compromise the information had relatively little effect on willingness to install in our first survey (see Section~\ref{sec:results:s1privacy}), we consider only proximity-based data in this scenario. Future work may wish to replicate these results for location information.
\begin{quote}
    Please consider the following scenario. Imagine that public health workers will notify you if they are able to determine that you have recently been near (within 6 feet) someone who was diagnosed with coronavirus.
    \begin{itemize}
        \item You do not have to do anything in order for the public health workers to monitor whether you have recently been near someone diagnosed with coronavirus.
        \item However, the public health workers are not aware of every time you are near someone diagnosed with coronavirus.\\
    \end{itemize}

Imagine that there is also a mobile phone app available that will alert you if you have been within 6 feet of someone diagnosed with coronavirus.
\begin{itemize}
    \item The app will do this by collecting information about who you have been within 6 feet of (who you have been "near").
    \item The app will not reveal the identity of the people you have been near.
\end{itemize}
\end{quote}

Participants were then assigned to one of the branches in Table~\ref{tab:rq3conditions}.
\begin{table}
\centering
\scriptsize
\begin{tabular}{llll}
    Condition & Questions Asked & Without-App Baseline & Variable\\
    \hline
    Public benefit & Implicit privacy assessment, public benefit & 3\% IHME infection rate~\cite{Infectio80:online} & Infection reduction (97\%, 83\%, 66\%, 50\%, 33\%, 10\%)\\
    False negative & Implicit privacy assessment, FN & 1 in 100 exposures detected & FN rate (90\%, 50\%, 10\%) + control (99\%) \\
    False positive & Implicit privacy assessment, FP & 1 in 100 exposures detected without app, all with app & FP rate (1\%, 10\%, 25\%, 50\%)\\
    Privacy & Privacy, FN & -- &  Risk of compromise (1 in 10, 1 in 100, 1 in 1000) \& FN as above\\
    \hline
\end{tabular}
\label{tab:rq3conditions}
\caption{Survey branches in RQ3 survey}
\end{table}

No information in this survey was expressed in terms of percentages, due to a plethora of research in health risk and numeracy showing that people interpret rates far more accurately than percentages~\cite{riederer2018put,keller2009effect}. Below we describe exactly how each of the questions referenced in Table~\ref{tab:rq3conditions} was asked in our survey.

\textit{Implicit privacy.} Pilot tests of our survey revealed that people had implicit privacy perceptions of the app described, which were influencing their willingness to adopt the apps. We used a modified version of the Paling Perspective scale~\cite{paling2003strategies} -- a well validated tool for eliciting health risk perception -- to assess respondents' perception of the likelihood that information collected by this app would be compromised~\footnote{This digital safety risk-perception assessment technique was developed and validated by Cormac Herley, Elissa M. Redmiles, and Siddharth Suri; paper under review.}. This measurement allows us to (a) report on people's perceptions of the likelihood that information from a coronavirus app will be compromised, (b) control for the effect of differing implicit privacy perceptions on willingness to install, and (c) validate the influence of these privacy perceptions by comparing willingness to install given an implicit privacy perception vs. an explicit one that we set by telling the participant the risk their privacy will be compromised (described in the next section). 

The question we used to assess implicit privacy belief was:
\begin{quote}
    Studies show that despite best attempts to protect the data of those who use this app, some people may have information about who they have been near compromised and used for purposes other than the fight against coronavirus.\\

Please indicate on the chart below how many app users you think will have this information compromised over the next year.
    \begin{figure}[h!]
        \centering
        \includegraphics[width=0.7\textwidth]{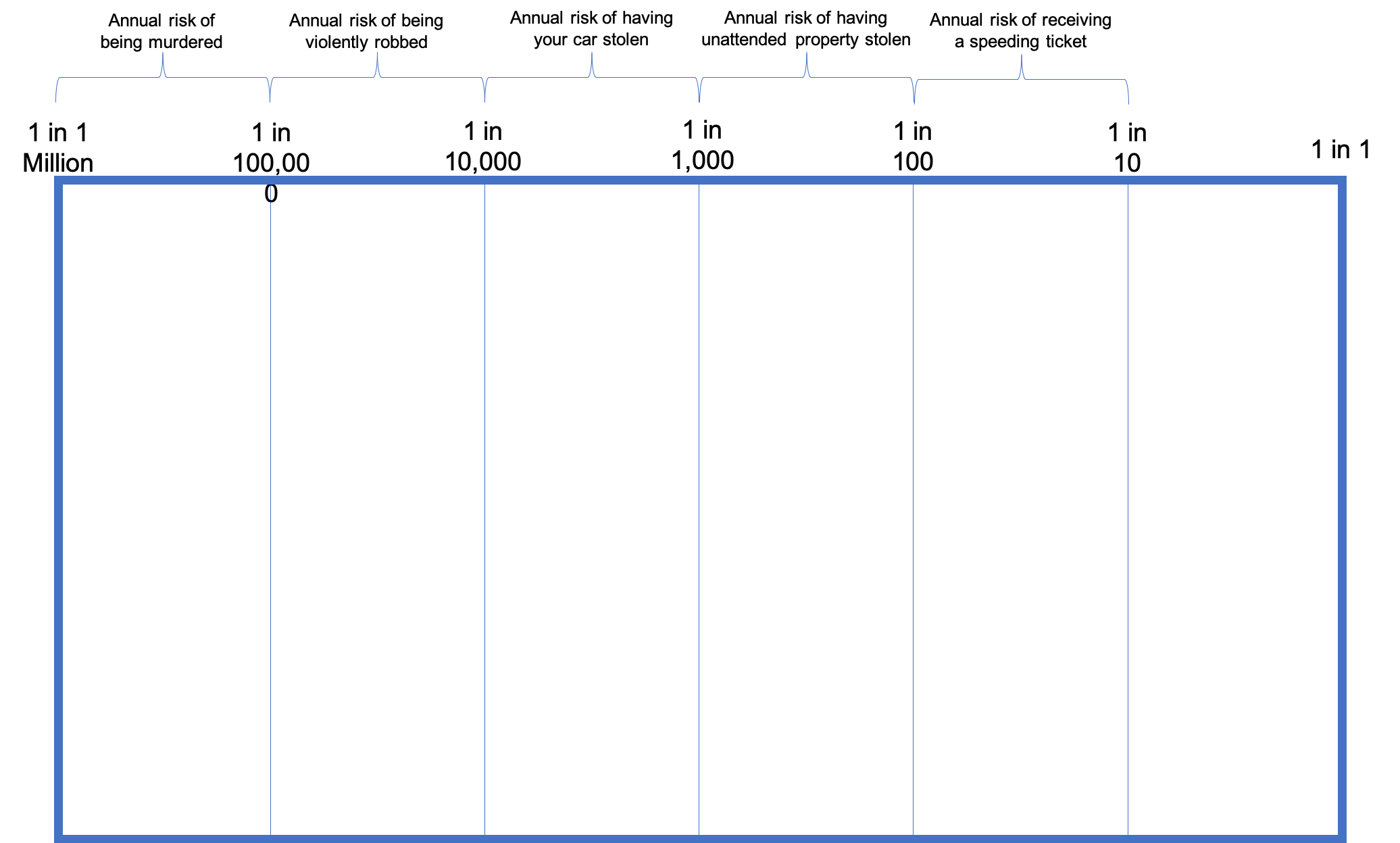}
    \end{figure}
\end{quote}

\textit{Explicit Privacy.} In order to understand how different privacy risks impacted respondents' reported willingness to install coronavirus apps we asked some participants about their willingness to install in the context of explicitly known (as opposed to implicitly perceived, as aforementioned) privacy risks. We asked about explicit risk using the following question:
\begin{quote}
    Studies show that despite best attempts to protect the data of those who use this app, some people may have information about who they have been near compromised and used for purposes other than the fight against coronavirus.\\ 
    
    \textit{X} out of 1000 people who use this app will have this information compromised.
\end{quote}

We also asked all participants in this branch the false negative question (below) in order to be able to cross-validate the impact of explicit declaration of privacy risk vs. the effect of implicit perception of that risk on willingness to install. As it would not make much sense to assess implicit risk and then ask respondents whether they would install given their implicit perception of risk, privacy questions need to be paired with a benefit question. We chose to make our comparison using false negative questions since the results of survey 1 showed that false negatives were equally as important in users' consideration of whether to install as was privacy.

\textit{False negative.} We asked respondents whether they would be willing to install an app that could detect N out of 100 exposures to coronavirus compared to manual contact tracing, which could detect exposures a baseline number of times: 1 out of 100. The question was phrased as follows: 
\begin{quote}
    Imagine that you are exposed to someone who has coronavirus 100 times over the next year.\\

\textbf{If you do not use the app}, 1 out of 100 times public health workers will be able to detect and notify you that you were exposed.\\

If you use the app, \textit{FN} out of 100 times the app will be able to detect and notify that you that you were exposed.
\end{quote}

To compare willingness to install a 1\% effective app as a baseline, we also had a FN control condition. This condition consisted of a scenario that did not describe manual contact tracing, but just described the app (in the same way as above), which respondents were told could detect 1 in 100 exposures (the same as the manual contact tracing option offered in the other conditions).

\textit{False positive.} We asked respondents whether they would be willing to install an app that detected all exposures to coronavirus\footnote{We examined an app with 0\% FN as an upper bound on respondents' tolerance to FP.}, but had N out of 100 additional false negatives. The question was phrased as follows: 
\begin{quote}
    Imagine that you are exposed to someone who has coronavirus 100 times over the next year.\\

\textbf{If you do not use the app}, 1 out of 100 times public health workers will be able to detect and notify you that you were exposed.\\

The app is not perfect. \textbf{If you use the app}, the app will correctly notify you every time that you were exposed (100 out of 100 times). The app will also incorrectly notify you an additional FP times, when you were not actually exposed.
\end{quote}

\textit{Public Health Benefit.} Finally, some respondents were assigned to a branch that evaluated how reduction in infection rate among app users would influence people's willingness to install an app.

We chose 3\% as the baseline infection rate without app use as this is the currently estimated U.S. infection rate by the IHME~\cite{Infectio80:online}. 

\begin{quote}
    Studies show that 30 out of 1000 people who do not use the app will be infected with coronavirus in the next year.\\
    
    \textit{H} out of 1000 people who use the app will be infected with coronavirus in the next year.\\
\end{quote}

\textbf{Validation.} The questionnaire design was validated through expert reviews with multiple external researchers. Additionally, three attention check questions were included, one general and two scenario-specific, as in Survey 1.

\textbf{Sample.} 3,826 Amazon Mechanical Turk workers responded to our survey. These workers were split into different survey branches, as aforementioned, so all results sections note the number of respondents used in a particular analysis.

\textbf{Analysis.} We analyze the data obtained in this survey descriptively, through data visualization, and using binomial logistic regression analysis: with willingness to install as the dependent variable and the dependent variables of the varied factor (e.g., chance of FN) and perceived implicit privacy risk. To evaluate the impact of privacy on decision making we use a $X^2$ proportion test to compare the proportion of respondents willing to install given some FN rate in the implicit and explicit privacy conditions.

\subsection{Limitations} As with all surveys, the answers represented in these results are people's self-reported intentions regarding how to behave. As shown in prior literature on security, these intentions are likely to align directionally with actual behavior, but are likely to over-estimate actual behavior~\cite{redmiles2018asking}. The goal of this work is to show \textit{how} willingness to adopt may be influeced by privacy/accuracy considerations, and thus the precise numeric estimates should not be interpreted as precise adoption estimates. 

Additionally, regarding the RQ3 survey, there are always concerns about the generalizability of crowdsourced results. To address these concerns, we also conducted the RQ1,RQ2 survey on Amazon Mechanical Turk. We found only one significant difference (with small effect size) in the AMT results as compared to the online survey panel results. Due to the quantitative nature of the RQ3 survey and the sample size required, and our comfort in the relatively representative nature of AMT results on this particular topic verified by our RQ1, RQ2 comparison as well as prior work on the generalizability of AMT results in security and privacy surveys~\cite{redmiles2019well}, we chose to proceed with AMT for RQ3.

\section{Results}
In this section, we detail our findings. For those who prefer a swifter visual summary, please see \url{http://www.cs.umd.edu/~eredmiles/how-good-good-enough.pdf}. 
\begin{figure}[t]
    \centering
    \includegraphics[width=0.9\textwidth]{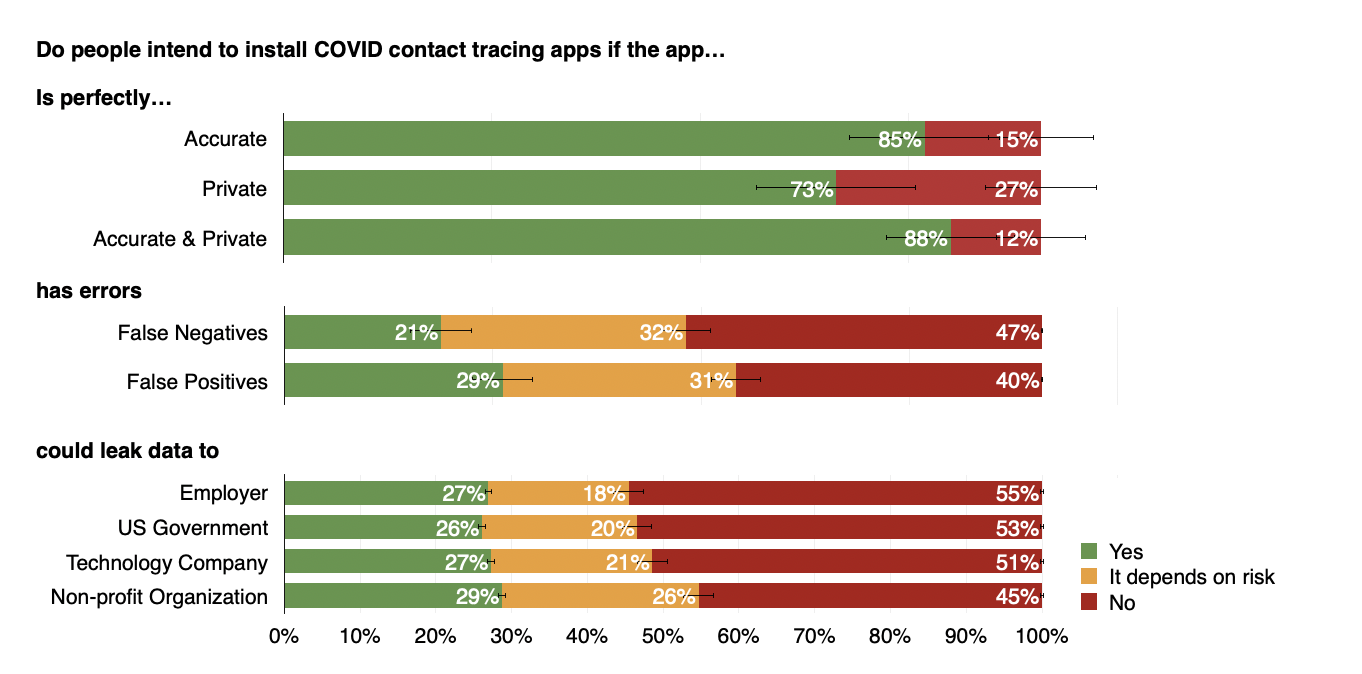}
    \caption{Results of a survey of the general population (n=789) on willingness to install a COVID19 contact tracing app. Std. error bars are shown.}
    \label{fig:s1results}
\end{figure}
\begin{figure}[t]
\centering
    \includegraphics[width=0.5\textwidth]{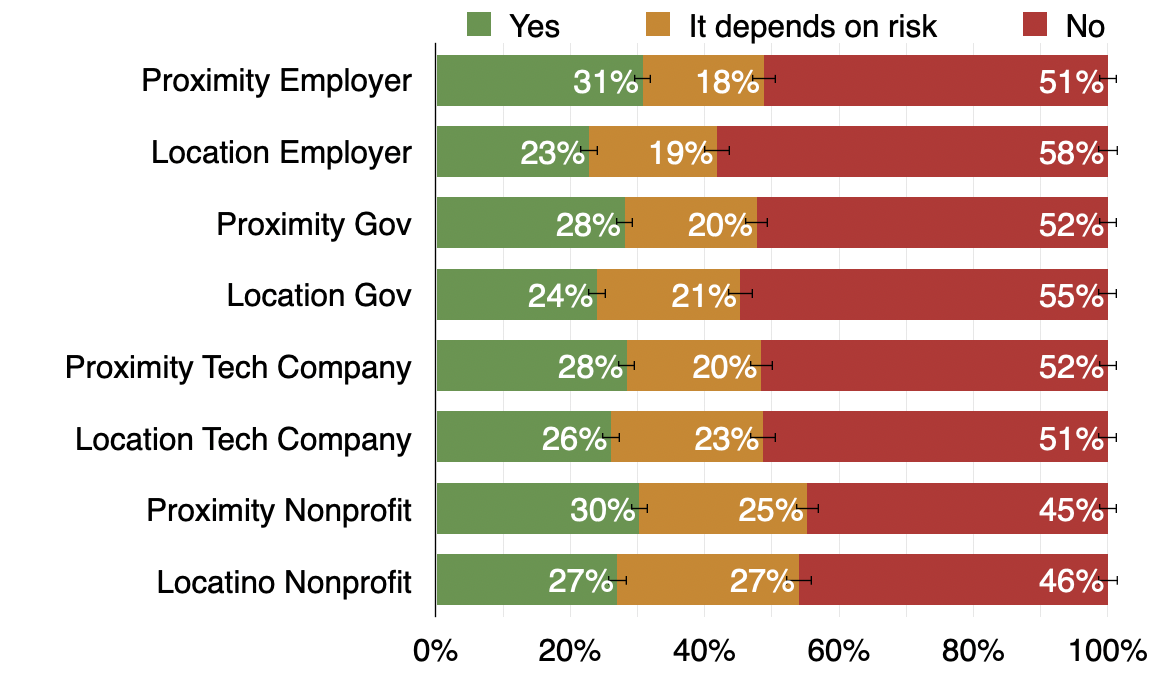}
    \caption{Reported willingness to install a COVID19 contact tracing app depending on the data it might leak (n=789, general population). Std. error bars are shown.}
    \label{fig:s1results-location-proximity}
    \end{figure}

\subsection{Both accuracy \& privacy influence whether people want to install a COVID app}

The results of our first survey, shown in Figure~\ref{fig:s1results} illustrate that both accuracy and privacy do indeed significantly ($X^2$ tests in comparison to the control conditions, $p<0.05$, Bonferroni-Holm multiple testing correction (BH correction)). 

We find that respondents did not significantly differentiate between perfect privacy vs. perfect accuracy ($X^2$ prop. test, p=0.178,  BH correction), perfect accuracy vs. both perfect accuracy and privacy ($X^2$ prop. test, p=0.670,  BH correction), of perfect privacy vs. both perfect accuracy and privacy ($X^2$ prop. test, p=0.069,  BH correction).

On the other hand, respondents were 8\% less likely to install an app with false negatives, regardless of the number of the FN rate, than one with false positives, regardless of the FP rate ($X^2$ prop. test, p=0.003,  BH correction). Respondents were similarly less likely to install an app with false negatives, regardless of the FN rate, than one with privacy leaks to any of the entities examined ($X^2$ prop. test, p=0.003,  BH correction). Respondents were equally as likely to install an app with false positives as one with privacy leaks ($X^2$ prop. test, p=0.609,  BH correction). 

Respondents were more likely to say that their decision to install would depend on the risk of false positives (31\%) or false negatives (32\%) than the risk of a privacy leak (17\% across entities).

Finally, respondents' reported willingness to install did not significantly differ ($X^2$ prop. tests, p>0.05, BH correction) based on what data the app might leak to a particular entity, except for hypothetical leaks to the respondents employer (Figure~\ref{fig:s1results-location-proximity}). Only 23\% of respondents were willing to install an app that might leak their locations to their employer while 31\% were willing to install an app that might leak information about who they have been near (their proximity data) to their employer. The next section provides regression comparisons of willingness to install based on the entity to which the information was leaked, and also control for data type differences (finding no significant differences).

\subsection{Some Americans weigh accuracy or privacy considerations more highly than others}
In order to examine whether some American's weighed accuracy or privacy considerations more highly than others, we constructed two mixed effects logistic regression models as described in Section~\ref{sec:s1methods}. 

\begin{table}[ht]
\centering
\small
\begin{tabular}{lrll}
 Variable & Odds Ratio & CI & p value \\ 
  \hline
 Question: False Positive & 1.65 & [1.18, 2.32] & $<$ 0.01** \\ 
Data: proximity & 0.94 & [0.55, 1.61] & 0.83 \\ 
Age & 0.99 & [0.97, 1] & 0.1 \\ 
Gender: female & 1.27 & [0.2, 8.21] & 0.8 \\ 
Know someone who died from COVID19 & 5.56 & [2.35, 13.13] & $<$ 0.01*** \\ 
High medical risk & 1.60 & [0.84, 3.03] & 0.15 \\ 
Internet Skill & 1.77 & [1.22, 2.56] & $<$ 0.01** \\ 
Pol. leaning: Democrat & 0.60 & [0.34, 1.05] & 0.07 \\ 
Edu.: BA+ & 1.24 & [0.61, 2.51] & 0.55 \\ 
Edu.: SC & 1.54 & [0.7, 3.36] & 0.28 \\ 
   \hline
\end{tabular}
\label{tab:regression:s1accuracy}
\caption{Mixed effects logistic regression model of willingness to install apps with accuracy errors. Question baseline is FN, data baseline is location, political leaning baseline is Republican, mixed effects term controls for within-subjects design.}
\end{table}
\begin{table}
\centering
\small
\begin{tabular}{lrll}
Variable & Odds\_Ratio & CI & p\_value \\ 
  \hline
Entity: Tech Company & 1.18 & [0.79, 1.76] & 0.41 \\ 
Entity: Employer & 0.81 & [0.54, 1.2] & 0.29 \\ 
Entity: NonProfit & 2.38 & [1.6, 3.56] & $<$ 0.01*** \\ 
Data: proximity & 1.97 & [0.92, 4.21] & 0.08 \\ 
Age & 0.93 & [0.91, 0.95] & $<$ 0.01*** \\ 
Gender: female & 0.38 & [0.17, 0.82] & 0.01* \\ 
Know someone who died from COVID19 & 1.14 & [0.37, 3.52] & 0.82 \\ 
High medical risk & 1.95 & [0.79, 4.8] & 0.15 \\ 
Internet Skill & 1.78 & [1.23, 2.58] & $<$ 0.01** \\ 
Pol. Leaning: Democrat & 2.70 & [1.23, 5.94] & 0.01* \\ 
Edu.: BS+ & 0.55 & [0.2, 1.51] & 0.25 \\ 
Edu.: SC & 0.82 & [0.27, 2.47] & 0.72 \\ 
   \hline
\end{tabular}
\label{tab:regression:s1privacy}
\caption{Mixed effects logistic regression model of willingness to install apps with privacy errors. See Table~\ref{tab:regression:s1accuracy} caption for more information.}
\end{table}

We find that those who know someone who died from COVID19 are over 5$\times$ as likely as those who do not to be willing to install an app that has errors in accuracy. Additionally, we validate that even when controlling for demographic variance, respondents are more comfortable with false positives than false negatives: respondents are 65\% more likely to report that they would install an app with false positives than one with false negatives.

Respondents were more comfortable installing an app with potential privacy leaks to a non-profit organization verified by the government an app with potential leaks to any other entity (their employer, a technology company, or the U.S. government). Those who identify as Democrats are nearly $3\times$ as likely as those who identify as Republican to be willing to install an app with privacy risks. Finally, those who are younger and women are less likely to report that they would install an app with privacy errors. This gender finding aligns with past work showing that women may be more privacy sensitive than men~\cite{redmiles2018net,hoy2010gender}. 

Finally, those who have higher internet skill are more willing to install an app that has either errors in accuracy or privacy leaks, likely because those with higher skill are more likely to install COVID19 apps in general~\cite{WillAmer35:online}. 

\subsection{\textit{Amount} of public health and individual benefit influence willingness to install}
\begin{figure}[t]
    \centering
    \includegraphics[width=0.6\textwidth]{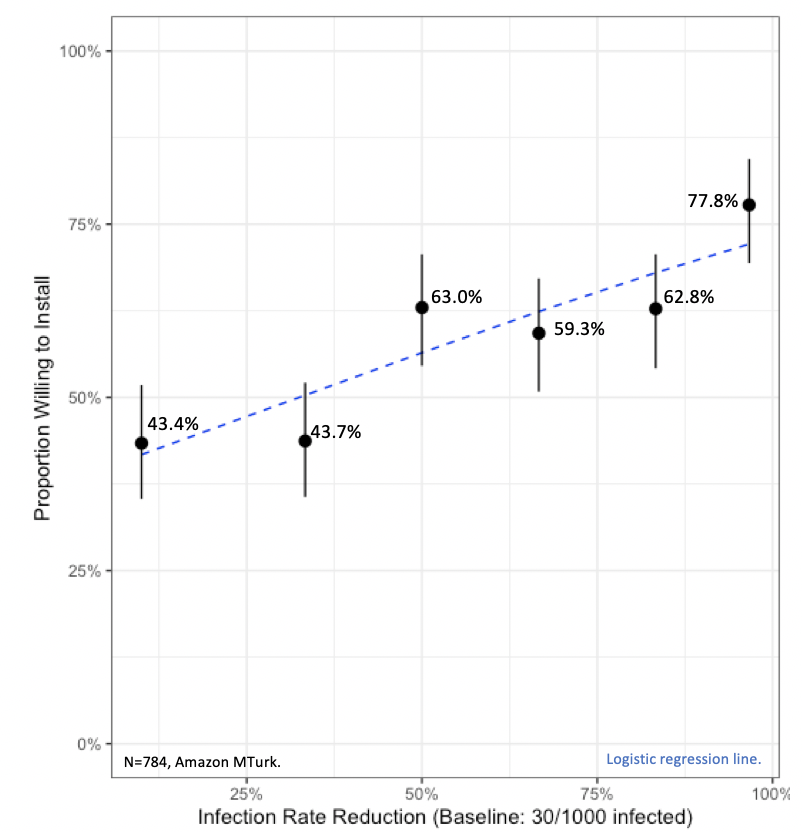}
    \vfill
     \includegraphics[width=0.45\textwidth]{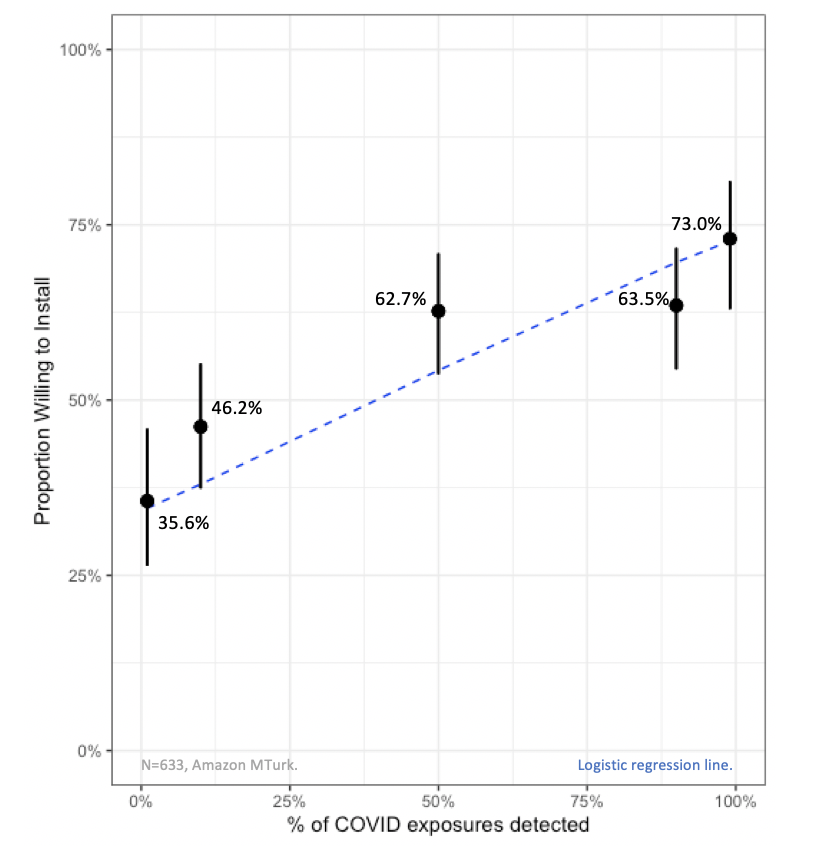} \hfill \includegraphics[width=0.45\textwidth]{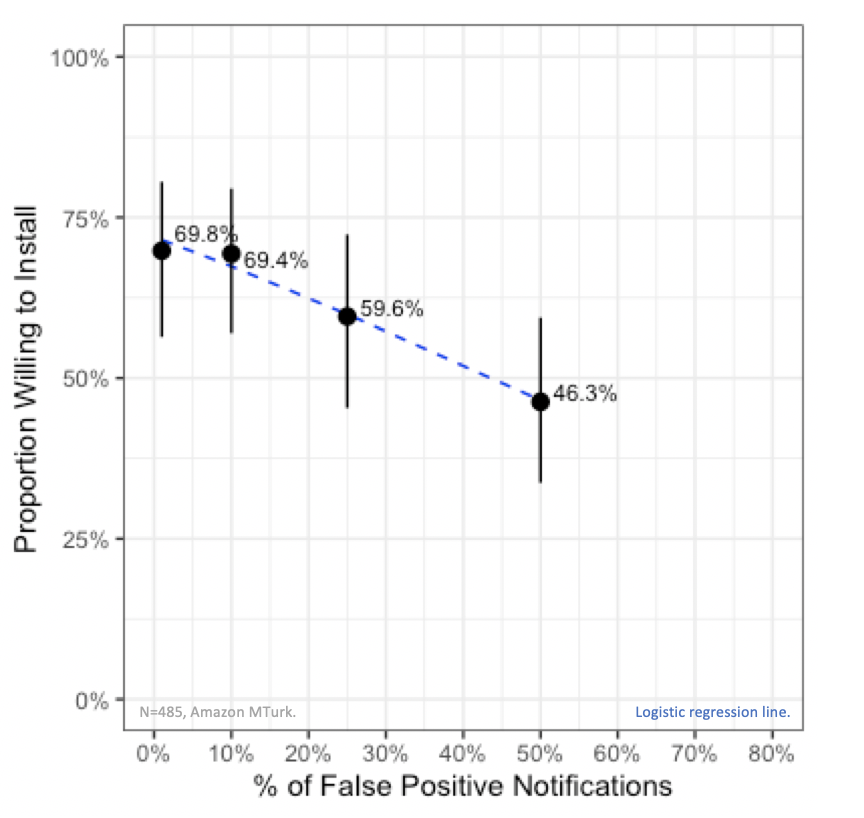}
    \caption{Relationship between willingness to install and amount of public health benefit (top) and individual benefit (bottom).}
    \label{fig:isolines}
\end{figure}

In the findings above, we validate that the individual considerations of accuracy and privacy both impact reported willingness to install. In our second survey we examine whether we can model how the quantitative \textit{amount} of public health (i.e., infection rate reduction) and individual benefit (i.e., FN and FP rates) influences willingness to install. Figure~\ref{fig:isolines} provides an overview of these findings.

%

To examine the relationship between amount of benefit and willingness to install beyond visual inspection, we construct logistic regression models. We find that, for every 1\% reduction to infection rate offered by the app, respondents are 5\% more likely to report that they would install (O.R., 95\% CI: [0.94, 0.97], $p<0.001$) 

Individual benefit has a relatively smaller effect on willingness to install. For every 1\% increase (O.R. 95\% CI: [1.01, 1.02], $p<0.001$) in app sensitivity (1\% decrease in false negative rate) respondents are 1\% more likely to report that they would install. Finally, for every 1\% decrease (O.R. 95\% CI: [0.98,0.99], $p<0.001$) in false positive rate, respondents are 1\% more likely to report that they would install.  

\subsection{Implicitly perceived risk of privacy leak in COVID apps influences willingness to adopt; risk of COVID app privacy leak perceived by respondents as between 0.01\% - 0.001\%}
\label{sec:results:s2privacy}

In our second survey, we not only measured willingness to install based on amount of benefit but we also measured implicit privacy risk perception. In this section we briefly summarize how likely respondents thought it was that information from a COVID19 contact tracing app would be leaked and we confirm the results of survey one: that privacy risk, even when unmentioned, influences willingness to adopt a COVID19 app. 

\begin{figure}
    \centering
    \includegraphics[width=0.9\textwidth]{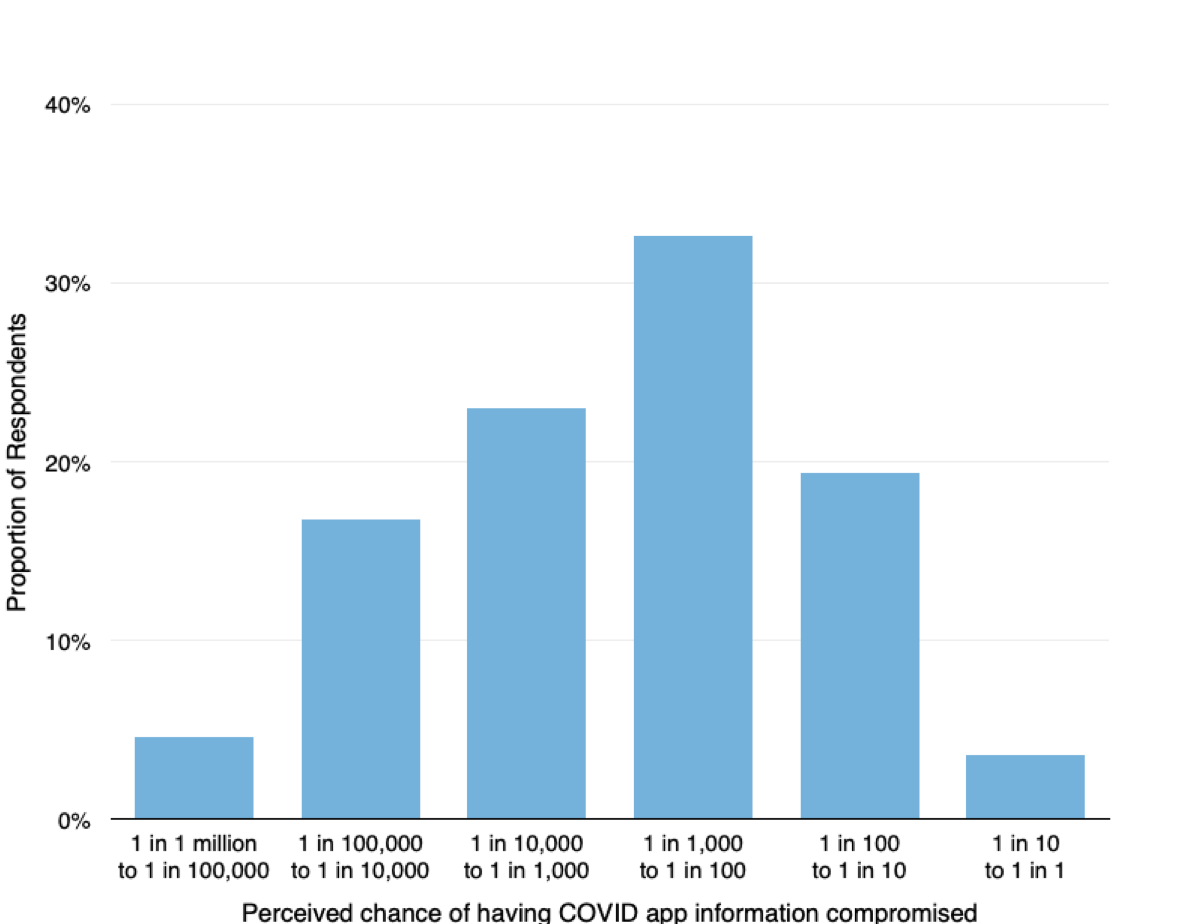}
    \caption{Respondents implicit perceptions of the chance that information from a COVID app would be compromised in the next year.}
    \label{fig:privacyrisk}
\end{figure}

Figure~\ref{fig:privacyrisk} summarizes respondents implicit perceptions of the risk of a privacy leak of COVID19 app information in the next year. The median respondent (n=1,610) reported perceived the risk of a privacy leak of app information in the next year as between 0.01 and 0.001\%, equivalent to the annual risk of an American having unattended property stolen.  82\% of respondents reported perceiving the risk as between 0.1\% and 0.00001\%.

Finally, we compare the proportion of respondents who were willing to install a COVID19 app given an explicit statement of privacy risk (privacy risks were drawn from the portion of the implicit risk distribution reported by the majority of respondents) vs. their own implicit perception. We find no significant different between the proportion of respondents who were willing to install an app with a given false negative rate when relying on their own implicit privacy assumption and the proportion who were willing to install given an explicit statement of the risk of privacy leak. 

A regression model of willingness to install in the explicit condition finds a significant relationship between risk perception and willingness to install (O.R.: 1.80, 95\% CI: [1.31, 2.44], $p<0.001$). This lends support for our implicit privacy risk measurements and suggests that these implicit risk perceptions affect willingness to install similarly to explicit risk statements. 

Finally, further confirming the relevance of all three components studied in this paper -- benefits, accuracy, and privacy -- in users' consideration of whether to install, when we add implicit privacy risk as a factor to the regression models for willingness to install dependent on public health and individual benefit, we find that it is significant in all three models.

\section*{Acknowledgements}
With thanks to Eric Horvitz for the idea of investigating quantiative tradeoffs in public benefit, accuracy, and privacy. With thanks to Cormac Herley and Carmela Troncoso for survey feedback and general contact tracing conversations that contributed to this paper.

This work was funded by Microsoft Research.

\bibliographystyle{acm}
\bibliography{main}

\end{document}